# An Optimize-Aware Target Tracking Method with Combine MAC layer and Active Nodes in Wireless Sensor Networks

Amir Javadpour[1,*]

[1]School of Computer Science and Technology, Guangzhou University, Guangzhou, China, 510006

*Correspondence to: a_javadpour@gzhu.edu.cn

*Abstract*— **Wireless Sensor Network (WSN) is consisted of nodes with different sizes and a specific goal. Tracking applications are very important in WSNs. This study proposes a method for reducing energy consumption in WSNs, considering target tracking. A prediction method is also presented with the help of MAC (Medium Access Control) layer in which a few number of nodes are in active mode to track the target while the rest of the nodes are switched into sleeping mode. Energy consumption was reduced by decreasing the number of nodes involved in target tracking and activating merely a limited number of necessary nodes. The conducted experiments suggest that the proposed algorithm has a much better performance compared to similar algorithms in different conditions, including energy consumption in different communication ranges, effect of the number of nodes on energy consumption, the number of nodes involved in tracking in terms of communication range, and losing the target. This means the proposed method is superior and more efficient. When the target enters the simulation area, nodes of the network identify the target node through the proposed method (which is the optimum method) and track it as long as it is inside the simulation area. Performance criteria, including energy, other nodes, and throughput were used in simulation. The results showed that the proposed method has the optimum performance and reduces energy consumption of the network during tracking. The results of the simulation showed that the proposed method reduces energy consumption to an acceptable extent, compared to previous methods, in addition to increasing the accuracy of target tracking.**

## I. INTRODUCTION

Various solutions have been proposed for the long-standing problem of tracking mobile targets by means of radars, robots, and the Personal Communication Networks (PCN). Studies on distributed tracking started with Distributed Sensor Networks (DSN) program in Defense Advanced Research Projects Agency (DARPA) in early 80s. The project was initially defined as "Tracking with Large Numbers of Small and Inexpensive Sensors with Wireless Communication Capabilities." This seemed an ambitious program considering the technology level of the time and they virtually had to study networks with a small number of nodes, which were large sensors (such as radars). By the advances of the technology in recent years, the initial program defined for distributed sensor networks activated and today large numbers of small sensor nodes are applicable for distributed tracking as Wireless Sensor Network[1]–[4]. WSNs contain large numbers of wireless sensor nodes that can sense one or more physical phenomena. These networks have drawn much attention due to large range of various applications they cover. One such application is target tracking [5], which in turn, consists of a wide range of various applications, including tracking of the enemy, animals, human targets, cars on the highways, and many more. On the other hand, sensor nodes of WSNs usually use batteries as energy source, recharging and replacing of which is very costly and sometimes even impossible. Therefore, maximizing the life span of the network through reduction of energy consumption of each node and balancing energy consumption of all nodes is the main challenge that research on target tracking in WSNs are faced with [4], [6], [7] Different methods have been studies, considering various criteria such as scalability, communication overheads, energy consumption, and target tracking accuracy[7], [8] and [5]. Have classified target tracking methods into cluster-based methods, tree-based methods, Mobicast message-based methods, prediction-based methods, and hybrid methods. In cluster-based tracking algorithms, node members of the cluster identify the target and send the information to the cluster head. The cluster head collects all the information, calculates the location of the target and finally sends the information to the sink node (central receiver) [9], [10]. Given the cluster forming and calculation of the target location in each iteration of the algorithm, this method has a high time and calculation complexity and can be improved. Similarly, in tree-based algorithms, nodes form the tree structure before the network starts working or as soon as the target is identified and each node sends its information to the parent node. As the information is cumulated in the root node, the location of the target is identified and the corresponding report is sent to the sink node [11]–[14]. Similar to cluster-based method, this algorithm also requires much time and formation of leaves and transmission of information from leaves to parent and from parent to sink takes long periods of time. If the target is mobile, then the process will be more complex. Mobicast message-based methods depend on prediction. In these methods, a message is sent to a group of nodes that are changing based on the estimated speed of the target. This message contains the time and location of the discovered object. Transmission of a group message leads to traffic increase and as a result, the speed of target finding decreases[8]. Prediction-based algorithms predict the next location of the target based on its recent speed and direction. Consequently, nodes of other areas, which are mainly neighbor areas where the target is more likely to go, activate before the target reaches them and switch back to sleeping mode after the

`

target passes them. In these methods, sleeping mode is where the sensor is not tracking and sensing any targets and thus all its components are off except for the radio section of the sensor, which is in idle mode and awaits a signal from other nodes to start tracking and monitoring targets again [7]. Additionally, other nodes, which are not in the path of the target, are also kept in sleeping mode so that energy consumption is reduced and the network's life span is increased. A challenge prediction-based algorithms face is activation of large numbers of nodes during their implementation. That is because the predicted next location of the target can cover a large area with a large number of sensor nodes. However, given the direction of the movement of the target, activation of many of these nodes is not necessary[15], [16]. Hybrid algorithms use a combination of the above-mentioned methods. Hybrid tracking methods are those tracking algorithms that meet the requirements of more than one type of target tracking methods. Distributed Predictive Tracking (DPT) [17] and Hierarchical Prediction Strategy (HPS) [18] are two examples of such algorithms. Due to the fact that problems of the existing methods reduce the quality of the network and thus reduce the accuracy of target tracking, providing methods to increase the accuracy and optimize energy consumption is very important. This is what this paper is trying to do. The second section of this paper reviews previous works in the literature. The third section explains the proposed method and describes the problem. The fourth section evaluates the results of simulation. Finally, the fifth and last section presents conclusion and future works.

## II. LITERATURE REVIEW ON TARGET TRACKING

The [19] (2011) has used Rumor method for tracking and routing. This method is a moderate state between announcing the existence of new information in the sensor node and requesting the information to be sent from the destination to the sensor node. Therefore, it functions as an interface between two networks that use flooding routing method, where the request to send information is distributed in one network while announcement of receiving new information is distributed in the other.

The [20] (2011) has used Multi Sink method to track several objects. In this method, each node sends a copy of information to all the neighbor nodes in order to distribute information. Upon receiving the information, every node resends the information to its neighbor nodes, save for the node it has received the information from. The time needed for a node to receive information and send it to other nodes is called a Round. The algorithm ends when all nodes of the network have received a copy of the information. The [21] (2014) has used Energy-Efficient Routing Protocol (EERP) for tracking and routing. This method uses geographical information for target tracking. It sends requests, which often include geographical properties. This protocol aims to limit and exclude the number of interests in release and direct distribution and to assign a specific area to sending interests to the entire network. In GEAR, each node holds an estimated cost and a learning cost in case of reaching the destination through its neighbors. The estimated cost is a combination of the remaining energy and the distance from the destination. The learning cost is a refinement of the estimated cost that is calculated for routing around the holes of the network. A hole is created when the node has no neighbor closer to the target area than itself. The estimated and learning costs are equal where there is no hole. When a packet reaches the destination, it jumps back so that to arrange the launch and create path for the next packet. That is how the learning cost is calculated. The [11] has used method Energy-Aware target tracking by reducing active nodes in Wireless Sensor Networks. This paper proposes a new tracking algorithm reducing the number of active nodes in both positioning and tracking by predicting the target deployment area in the next time interval according to some factors including the previous location of the target, the current speed and acceleration of the target without reducing the tracking performance. There is not scheduling activates for transmuting data and all sensor nodes available in the target area by predicting the target position in the next time slot interval. This The [22] has used hybrid method of Geographic Adaptive Fidelity (GAF) for tracking and routing. GAF is an energy-aware position-based routing algorithm that is primarily designed for mobile networks, but may also be equally applicable to sensor networks. GAF appropriately preserves energy by shutting down unnecessary nodes of the network without any change in the level of routing. It creates a virtual grid for the covered area in which each node uses its position, identified with GPS, to integrate itself with a target. Nodes integrated with the same nodes are considered equivalent packets in terms of routing costs. This equivalence is applied to some nodes in a unique network environment in the sleeping mode to store energy correctly. Therefore, GAF can significantly increase the life span of the network as much as the number of nodes increases. The [23] (2012) has presented Hybrid Clustering for Multi-Target Tracking (HCMIT) in WSNs. In this paper, a dynamic clustering algorithm is presented for tracking distributed objects with the help of audio information. A hierarchical sensor network system is a combination of high-capacity and low-dispersion nodes with limited but dense power. High-performance nodes play as cluster heads and low-performance ones as cluster members. Cluster heads closer to the target are more likely to be activated than those far from the target are. Similarly, the probability that a cluster member sends data to the cluster head depends on its distance from the target. The [24] (2012) has used target tracking method based on data-driven protocols. In most target tracking applications of WSNs, the extended absolute number has made assigning global identifiers to each node impossible. Lack of a global identifier, along with the random expansion of sensor nodes, makes it more difficult to select a unique set of studied sensor nodes. Furthermore, sensor node data is transmitted with significant redundancy within one area. Therefore, it is ineffective considering energy consumption and routing protocols that will be able to select a set of sensor nodes and use the data density that is addressed during data recovery. This careful consideration leads to data-driven routing that is different from address-based routing. In data-driven routing, the base station sends the query to specified areas and waits for data to be received from sensors located in selected areas. This data-driven protocol addresses data transfer between nodes to remove additional data and save energy. Later, a direct release is created, which is an unexpected development in data-driven routing. Most other direct-release-base protocols are proposed based on a similar concept. The [25] (2012) has used Hybrid Energy-Efficient Distributed (HEED) clustering algorithm for tracking and routing. HEED is different from the previous algorithms in that it uses several parameters

`

(rather than one) to form its subnets, which was mentioned in the previous section. The objective here is to allow each node to use its lowest level of power (as much as possible) in transmitting the data and store higher levels of its power to communicate with cluster heads, as well as sending data to the base station. The [26] (2013) has used hierarchical protocols for tracking and routing. Nodes with different capabilities can be used in WSNs. Some nodes have higher levels of energy, higher computing power and more memory compared to other nodes. The protocols used in this kind of sensor networks prioritize nodes. The network is divided into subnets; each consists of subgroups of nodes inside. Using specific algorithms, each node should transfer its information to its subgroup. The subgroup then sends this information directly or as a multiplex to the base station. Hence, we have a multi-layer protocol. The [27] (2014) has used P-LEACH (Low-energy adaptive clustering hierarchy) protocol for tracking and routing. This is a LEACH-based protocol that uses timing for adaptive mobility and adaptive traffic. In this protocol, cluster heads assign time intervals based on traffic and mobility of sensors. Each time interval, thus, is given two owners: An Original Owner and an Alternative Owner. The protocol keeps and serves new mobile sensor nodes in the database table, whether free or idle time intervals exist. The [28] (2012) has used Support Vector Machines (SVM) algorithm for multi-target tracking. This algorithm is based on Flooding and Support Vector Machine methods but it selected one node out of neighbor nodes randomly and sends information only to that node. Moreover, the receiver node can send the information to the sender node again, albeit if that node is selected through the random selection. Table1 is illustrated summarize of different methods in target tracking.

TABLE 1. SUMMARIZE OF DIFFERENT METHODS IN TARGET TRACKING

| Protocol | Description | Disadvantages | Advantages | Simulator |
|---|---|---|---|---|
| Rumor | An interface between two networks that use flooding routing method, where the request to send information is distributed in one network while announcement of receiving new information is distributed in the other. | All nodes can send an agent message, which increases the spatial complexity. | It contains a table of all routes, which is sent to all nodes so that they update their routes. | MATLAB |
| Sink | In order to distribute information in the network, each node sends a copy of the information to all its neighbor nodes. Each node then resends the information it has received to all its neighbor nodes. | In the main flooding method, each node sends the data to its neighbors, regardless whether that node has already received the information. This causes internal explosion. | The nodes do not need initial information about the structure of the network and neighbor nodes. | MATLAB |
| EERP | This protocol limits and excludes the number of interests in release and direct distribution and assigns a specific area for sending interests to entire network. | If the packet is reached the destination area, it can be released through Recursive Geographic Forwarding or Restricted Flooding in that area. | A neighbor is selected to forward the packet based on learning cost function. This selection can be updated according to convergence of learning costs during packet delivery. | NS2 |
| GAF | An energy-aware position-based routing algorithm that is primarily designed for mobile networks. | This protocol has delay and relatively high spatial complexity. | Neighbors arrange their sleeping time so that they preserve routing in the system correctly. | NS2 |
| HCMIT | High-performance nodes play as cluster heads and low-performance ones as cluster members. Cluster heads closer to the target are more likely to be activated than those far from the target are. | This protocol uses estimation algorithm for finding the target. Given the mobility of the targets, the algorithm cannot show the predicted location accurately. | Low delay for finding targets. | NS2 |
| Data-Driven Protocols | The extended absolute number has made assigning global identifiers to each node impossible. | Computational Complexity | Optimal Energy Consumption | NS2 |
| Hybrid Sensory Clustering | Uses clustering and hybrid sensory method for tracking and routing. | This algorithm does not balance the load. | Cost Reduction | NS2 |
| Hierarchical Protocols | Uses hierarchical protocols for tracking and routing. | Extra Overhead | Expandable, Easy management of information transfer routs and nodes | NS2 |
| F-LEACH | Uses timing for adaptive mobility and traffic. Cluster heads assign time intervals based on traffic and mobility of sensors. | Extra Overhead | Increases the packet delivery ratio and decreases energy consumption. | NS2 |
| SVM | Uses flooding and support vector machine methods for target tracking. | A node can receive the same information from different routes, which causes energy loss. | The information overhead problem on neighbors of the tracked node is solved by SVM method. | MATLAB |

.

## III. PROPOSED METHOD

This section presents the proposed method for finding node's location and studies its tracking. Sensors with identifiers are used for implementing the tracking protocol. The aim of the protocol proposed in this chapter is to implement a routing method for tracking mobile objects in the network. Scalability and energy consumption reduction are criteria that are taken into account in the proposed algorithm. Desirable performance of a target tracking method depends on the following [11]:

`

- Detector nodes can identify the target location using locating methods and the locations of detector nodes.
- The route the target takes over time can be tracked through frequent reports.
- Given the limited energy of the nodes, minimum numbers of sensor nodes are active at each point of time.

The details of the proposition are as follows:

M is a mobile target entering the network at time k and point $(x_k, y_k)$ (Figure 1 (A))
i and j are the two sensor nodes closest to M at time k
i and j nodes can sense the target and their distances from M are $d_i$ and $d_j$, respectively
The speed of M is $V_k$ (Figure 1 (B))
Assuming that the speed of the target does not exceed $R_s/T$ during $[k, k+T]$ ($V_k \leq R_s/T$).

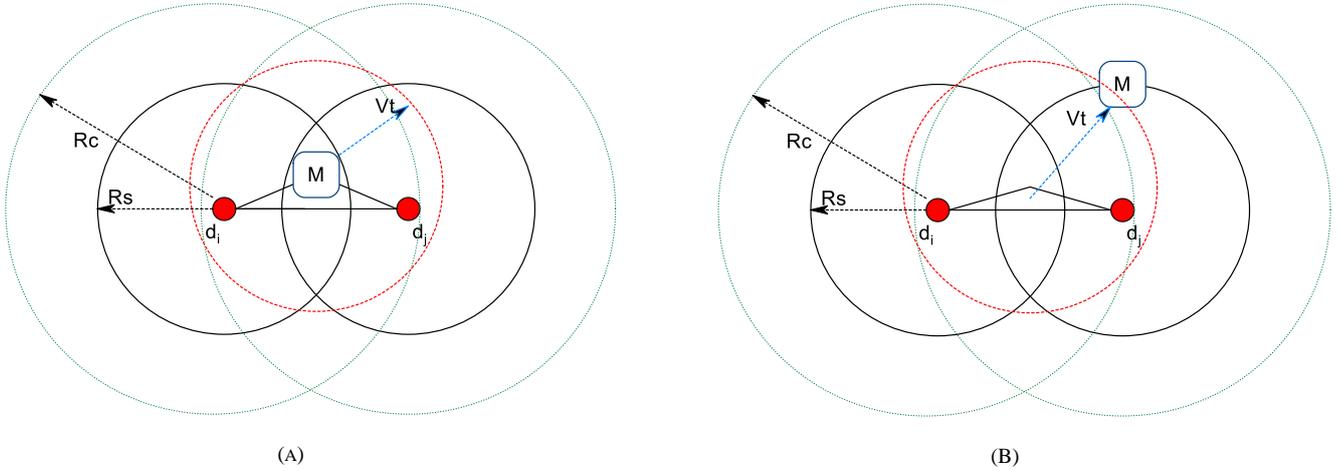

FIGURE 1 DETAILS OF PROPOSITION.
(a) i and j are the two sensor nodes closest to M at time k, (b)The speed of M is $V_k$

The proposition's overall view is as follows:

The target's location area at the next time interval is a circle centered on the current target coordinates and a radius proportional to the current target speed (Figure 2). A specific limit is set for the speed of the target, which is equal to $V_k \leq R_s/T$ based on proposition presented in [11].

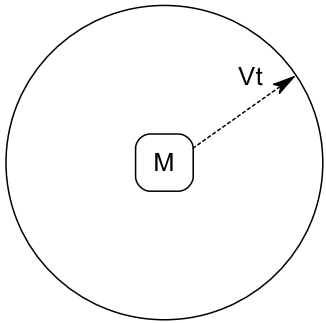
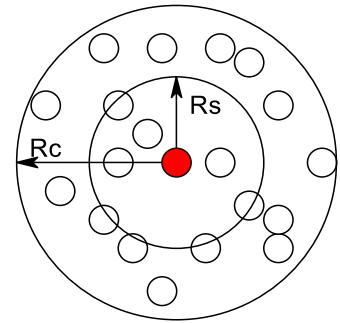

FIGURE 2 TARGET'S LOCATION AREA AT THE TIME INTERVAL.    FIGURE 3 COMMUNICATION RADIUS $R_C$ AND $R_S$.

*A: Problem Statement*

In the field of target tracking, we aim to track the target that enters an area of sensor nodes and we want the target to be accessible until it exits the area. The proposed method is based on prediction and thus it attempts to predict the next location of the target. This section explains the problem. When a target enters an area, the aim is to track it with the least amount of energy consumption. On the other hand, we need to ensure that the target is tracked. We are first going to explain the assumptions of the proposed method. As shown in Figure 3, $R_c$ is the communication radius of a sensor node, which is its communication range. The sensing radius, which is the range in which the node senses the environmental factors, is shown with $R_s$. It is assumed that $R_c \geq 2 \times R_s$. In the proposed method, if the nodes involved in target tracking can identify the target in the time slot, energy consumption is reduces. Obviously, reduction of the number of active nodes involved in tracking, as long as it does not damage the quality of tracking, leads to increase of life span of the network.

`

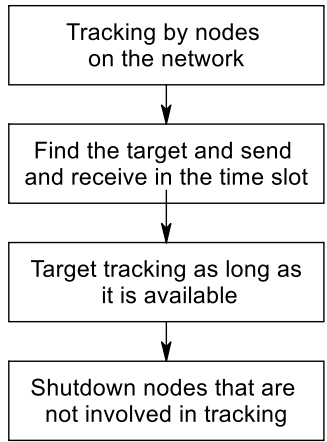

FIGURE 4 FINDING THE TARGET AND SENDING AND RECEIVING IN THE TIME SLOT

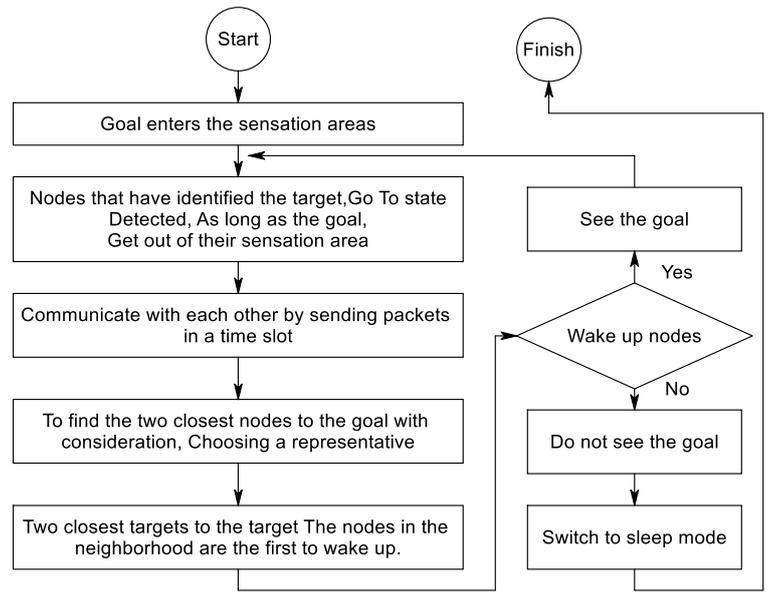

FIGURE 5 PROPOSED METHOD FLOWCHART

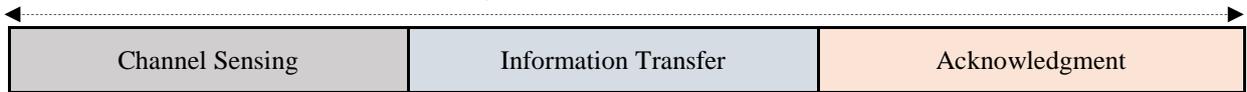

FIGURE 6 TIME SLOT STRUCTURE FOR NODES

*B: Proposed Algorithm*

In the proposed algorithm, nodes in the network are categorized into target detection, target monitoring, and off modes. When the target node enters the environment, other available nodes, based on their Rc and Rs being in the range of the target, sense it and change into detection mode. In order to identify the target and given the fact that the target is mobile, nodes that have sensed the target send waking messages to neighbor nodes. Upon receiving these messages, neighbor nodes are activated and switch into detection mode to identify the target. In our scenario, all nodes contain IDs (identification), which allow us to find out which nodes are involved in tracking. When the existing nodes observe the target in their radius, they notify each other of observation of the target through identification messages. The node with the least amount of numerical data is considered the representative node. The nodes that have not detected the target will switch into off mode in order to optimize energy consumption. Assuming that a slotted system is available, nodes involved in tracking use the same time slot model for sending and receiving data. At the beginning of each time slot the node with data to be sent is selected for sensing the channel and it is granted access to the channel based on the result of this sensing (Figure 4). Different algorithms are used for sensing and accessing the channel to maximize throughput, as well as reducing interference in order to reduce energy consumption. At the end of the time slot, the receiver notifies the sender of successful receipt of the information by means of ACK(). The time slot structure in the method proposed by this study is presented in Figure 5. As shown in flowchart in Figure 6, the target first enters the sensing area. Sensors that observe the target switch into detection mode and immediately start to communicate through sending packets. They cooperate with one another in this step in order to find the two nodes closest to the target and select a representative node. The representative node sends a message to the two closest nodes, which were identified and activated in the previous step, informing them that the target was observed. This message is sent in this step's time slot through metastasis and the algorithm moves to the next step after its acknowledgement. The acknowledgement validates transfer of the packet. The two nodes closest to the target wake their neighbor nodes. If no node observes the target, the representative sends no message and the two nodes closest to the target in the previous step realize that the target is lost. In this case, the newly activated nodes switch back to sleeping mode. However, if the target is observed, the tracking operation continues.

IV. EVALUATION

This section compares the proposed method with other algorithms. The most important parameter in WSNs is energy consumption. Here, we compare the proposed method with method presented in [11], namely, PBA (which was examined in the Background section). In PBA, all nodes are sensing and no pre-designed strategy exists. The simulation is conducted using NS2 software (Network Simulator2 [29], [30]) and output charts are created using MATLAB. Table 1 presents the simulation parameters and Table 2 contains network parameters of NS2. An example of the positioning of nodes in the simulation is also presented in Table 3.

TABLE 2.  TABLE 2 SIMULATION PARAMETERS[11].

| | |
|---|---|
| Initial Energy of the Node | 5 J |
| Sensing Radius (Rs) | 25 m |
| Communication Radius (Rc) | 50 m |
| Network Dimensions | 500 × 500 |
| Number of Nodes | 250 |
| Length of Control Packets | 512 |
| Mobility Model | Random Waypoint |
| Node Distribution Strategy | Random |
| Energy Consumption in Sleeping Mode | 0.00027 J |
| Energy Consumption in Sensing Mode | 0.012 J |
| Energy Consumption in Communicating Mode | 0.0378 J |

Sensing range radius 25 to 30 m

Communication range radius 50 to 60 m

Length of control packets 32 bite

Node initial energy 5 JE

amp 0.0013 9 0.000000000001 J

Eelect 50 9 0.000000001 J

TABLE 3.  TABLE 3 SIMULATION PARAMETERS OF NS2

| Network Interface Type | Phy/WirelessPhy |
|---|---|
| Channel Type | Channel/WirelessChannel |
| Propagation/TwoRayGround | Radio-Propagation Model |
| MAC Type | Mac/802_11 |
| Antenna Model | Antenna/OmniAntenna |
| Interface Queue Type | Queue/DropTail/PriQueue |
| Max Packet | Random Waypoint |
| Queue Type | LL |
| Number of Nodes | 250 |
| Max Packet | 512 |
| Routing Protocol | DSDV |

TABLE 4.  TABLE 4 EXAMPLE OF POSITIONING OF NODES IN THE SIMULATION AREA

| Node (0) | Set X_5.0 |
|---|---|
| Node (0) | Set Y_2.0 |
| Node (0) | Set Z_0.0 |
| Node (1) | Set X_390.0 |
| Node (1) | Set Y_385.0 |
| Node (1) | Set Z_0.0 |
| Node (2) | Set X_265.0 |
| Node (2) | Set Y_209.0 |
| Node (2) | Set Z_0.0 |

*A: Performance Criteria*

Throughput is calculated by Equation (1):

$$Throughput = \frac{\mu}{t} \qquad (1)$$

Where $\mu$ is the number of received bits and t is time (for the entire network, the result will be equal to average throughput of all node.)

End-to-End delay is calculated through Equation (2):

$$D = T_d - T_s \qquad (2)$$

Where D is the delay of one packet, Td is the receiving time of the packet at the destination and Ts is the sending time at the origin.

Packet delivery rate is calculated by Equation (3):

$$PDR = \frac{Receive\_Pckt}{Sent\_Pckt} \qquad (3)$$

Energy consumption of the node is calculated through Equation (4) (Energy Consumption Model):

$$NodeRemEng = NodeRemEng - (Etx \times Nt + Erx \times Nr + Eix + Esx) \qquad (4)$$

Where Etx is the energy consumption for sending the packet, Erx is the energy consumption for receiving the packet, Eix is the initial energy consumption, Esx is the cumulated energy consumption of all nodes, and N is the number of nodes. Figure 7 presents an example of target tracking simulation in a space with random distribution of nodes. The target rout is indicated with dashed line and the target stopping point is specified with a green star-like symbol. Sensor nodes in sleeping mode are specified with blue hollow circles, nodes that the target is located in their sensing radius are specified with ⊕ and nodes that switch into monitoring mode after activation are specified with ⊕. The target is examined as soon as it enters the simulation area and it is continued to be tracked until it exists the area.

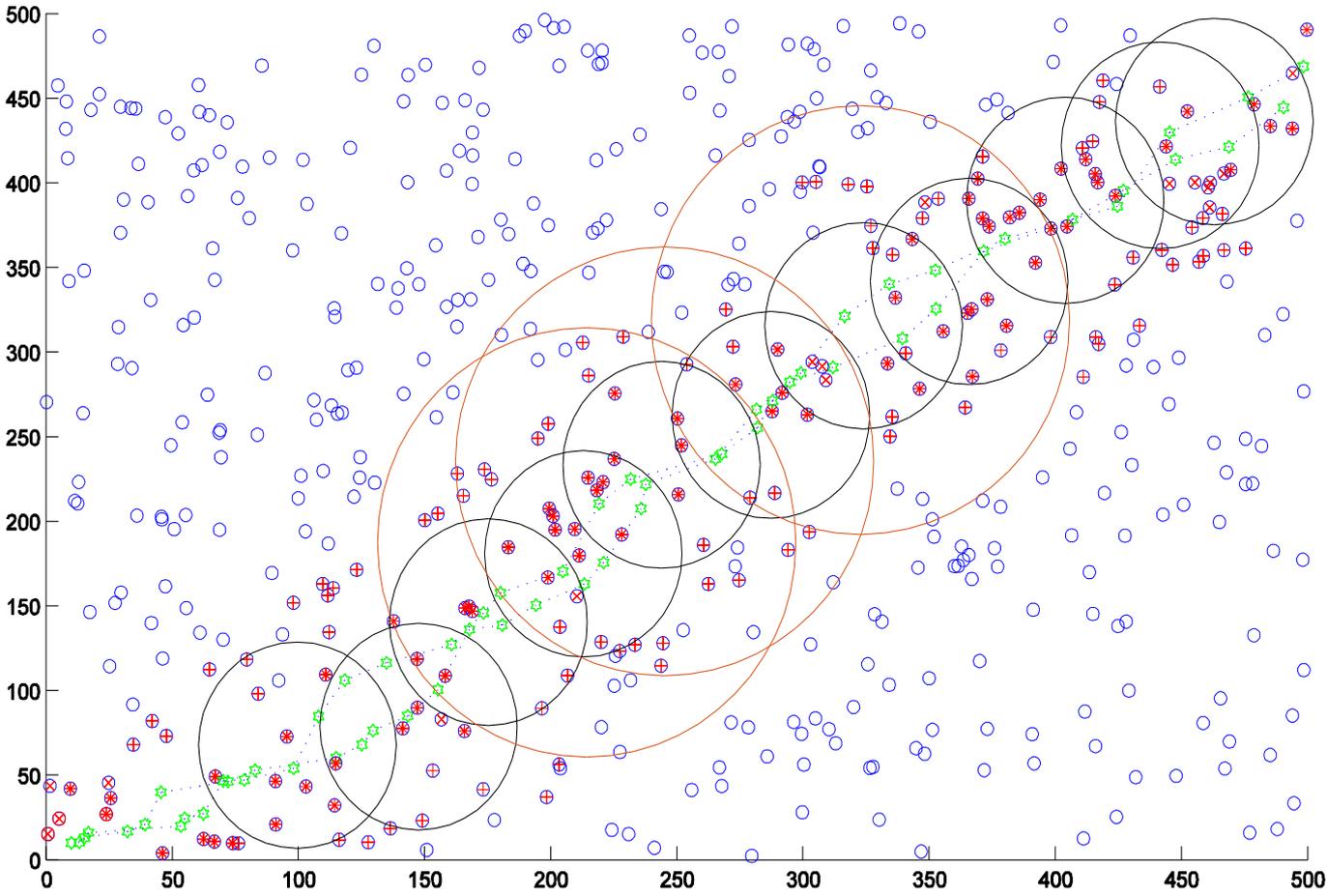

FIGURE 7 TARGET MOVEMENT IN A SPACE WITH RANDOM DISTRIBUTION OF NODES

*B: Analysis of Results*

The first experiment addressed energy consumption. This experiment examined the changes in sensing and communication radiuses of the existing nodes in the simulation (Figure 8). Energy consumption in the proposed method was optimized with the changes of communication radius compared to EATT method [11]. Considering that the information of the time slot was used for tracking the target node, repeated transmissions were avoided and eventually, the energy consumption was optimized. In the proposed method, the energy consumption with changes in communication radius was optimized by less involvement of target observing nodes. In this experiment (Figure 9), the energy consumption in the network was studies by changing the number of nodes in the simulation with a communication radius of 50 meters. By increasing the number of nodes in the communication radius areas the proposed method became optimum compared to EATT [11] and the energy consumption rate changed with a moderate slope. The experiment in Figure 10 shows examination of nodes involved in target tracking process. The number of nodes involved in target tracking is shown based on communication radius in the proposed method and in EATT. As the changes in communication radius increase, the number of involved nodes in the proposed method is less than EATT [11]. This indicates decrease of energy consumption in the network. The next experiment studies the energy consumption where the number of nodes in the network equals 250.

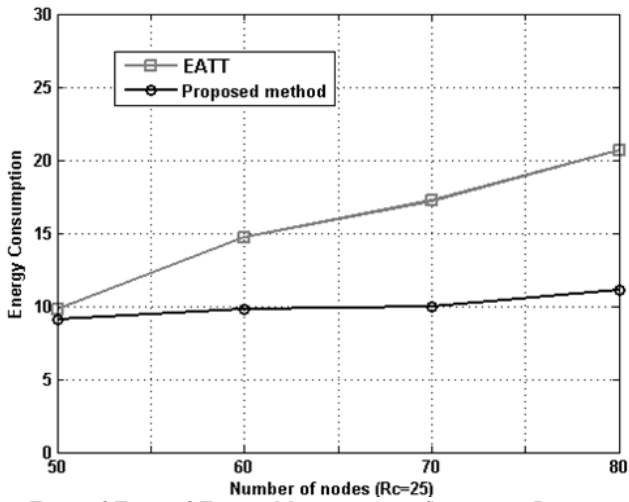
FIGURE 8 FIGURE 8 TARGET MOVEMENT IN A SPACE WITH RANDOM DISTRIBUTION OF NODES

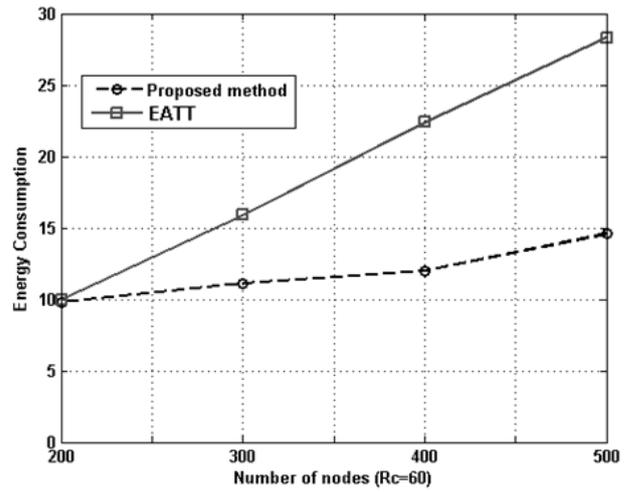
FIGURE 9 EFFECT IF NUMBER OF NODES ON ENERGY CONSUMPTION

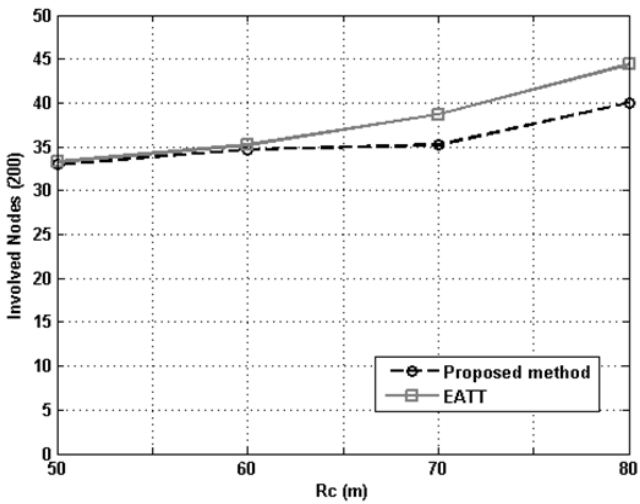
FIGURE 10 NODES INVOLVED IN TRACKING BY COMMUNICATION RADIUS AND 200 NODES

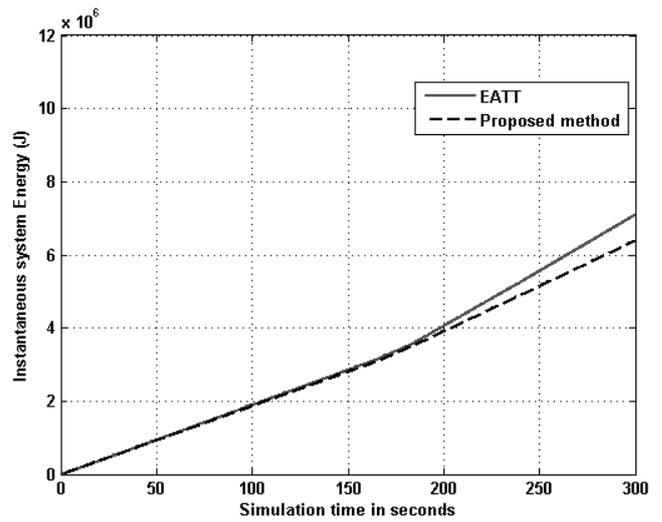
FIGURE 11 TOTAL CONSUMED ENERGY BY THE STEPS OF THE NETWORK

*C: Calculating Total Energy*

The energy consumed by the nodes consists of the energy used for saving the data and data recovery. Figure 11 shows a comparison between consumed energy by steps in the proposed method and EATT. When the network operates based on EATT method, a large amount of energy is lost due to nodes being permanently active and unnecessary listening during the idle time of the channel. The energy consumption of the proposed protocol is optimum considering its infrastructure. As shown in Figures 11, the energy consumption of the proposed method optimum by repeatedly sending the message on the route between the sender and the receiver.

*D: Calculating the Throughput*

A counter is set in the data entry section for the node parameter of NS2 in order to calculate the proposed method's throughput. This counter sends 3000 packets with specific length (byte) and different data rate (byte/s). This information moves between the specified origin and destination nodes. By each iteration of the proposed algorithm, sections related to Cyclic Redundancy Check (CRC) and the acknowledgment of the sent frame become inactive. The results show that changes in setting of data rate speed lead to different throughputs in the implementation system. The throughput changes by changing the data rate (data transfer speed) set in the sender node. As shown in Figures 12 to 15, the throughput of the proposed method increased compared to EATT by increasing the bandwidth rate and data transfer speed. The experiments were conducted based on the number of iterations (for x bytes) between the sender and the receiver. When the transfer rate is in 1 MB/s, the output of proposed method will have higher throughput. As the data rate is increase, the throughput of the proposed method is still higher than EATT. This indicates that the network remains stable and in good condition when the throughput increases. The proposed method was also efficient (compared to EATT) when the data rate increased in 3 and 4 MB/s.

`

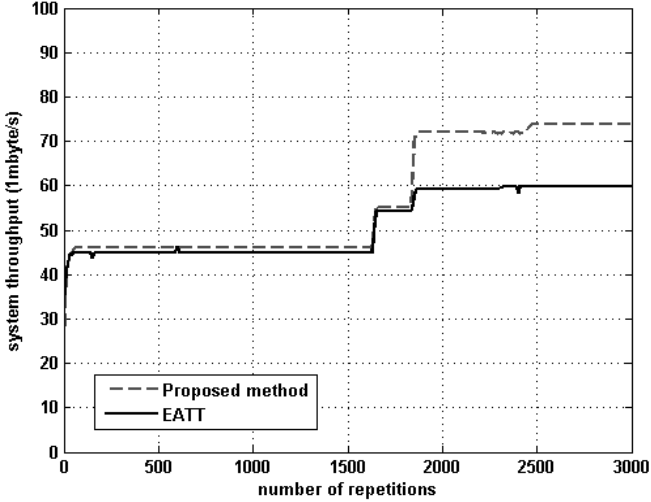
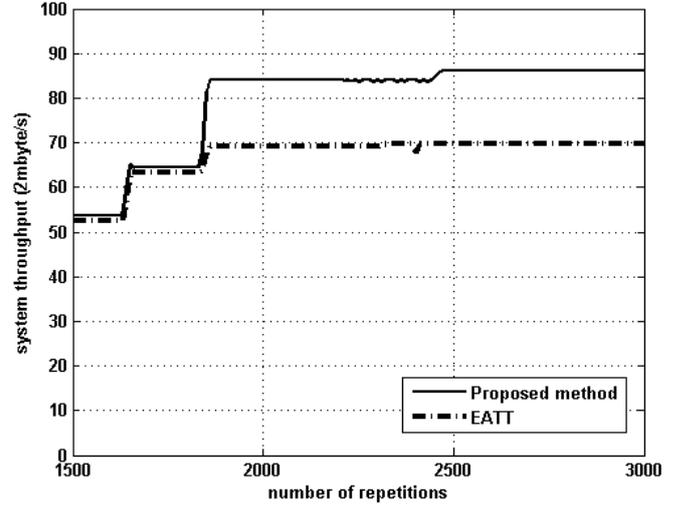
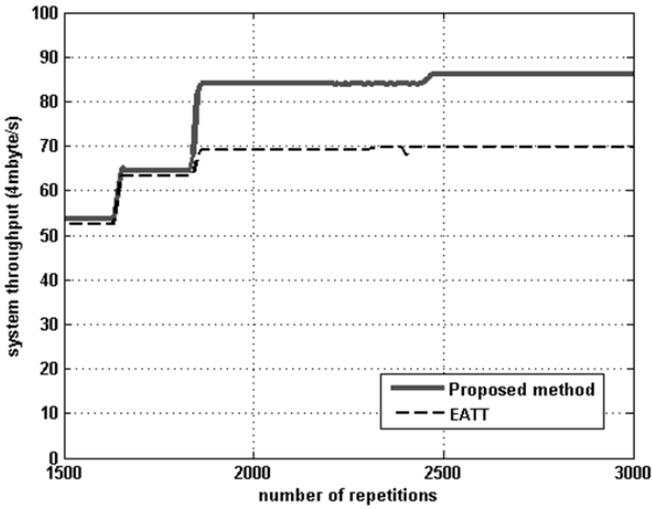
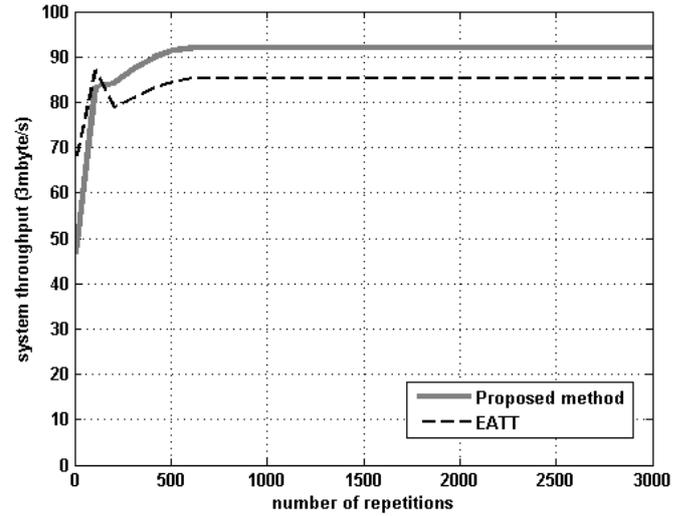

FIGURE 12 NETWORK THROUGHPUT (DATA RATES 1 TO 4 MB/S)

V. CONCLUSION

Wireless sensor networks consist of a large number of nodes with limited processing, wireless communication, and detection of one or more physical or chemical phenomena capabilities. The important feature of these nodes is the limited energy source that causes these networks to have a specific life span. Moreover, these networks are more limited compared to ad hoc networks in terms of processing and the range of radio communication. Therefore, algorithms proposed for these networks should be designed efficiently. On the other hand, these minimal features mean that the wireless sensor nodes are low-cost, making it possible to use a large number of these nodes in a network. Table 5 summarizes the considerations and criteria discussed and the relationship between factors that affect the design of the algorithm and their associated quality tracking criteria. As seen in the table, protocols of different layer encounter different quality tracking parameters. Although the information on the target mobility model reduces the scalability or flexibility of the tracking algorithm, it helps reduce energy consumption, which ultimately is a good quality-tracking criterion. This paper presents a prediction method using MAC layer, in which very few nodes are in the active mode for tracking while the rest of them switch into sleeping mode. The energy consumption was reduced be means of reducing the number of nodes involved in target tracking and keeping only a few necessary nodes for tracking the target. According to simulation results, the proposed method was able to reduce energy consumption, to an acceptable extent, compared to previous methods while increasing the tracking accuracy. The conducted experiments suggested that the proposed algorithm has a much better performance compared to similar algorithms in different conditions, including energy consumption in different communication ranges, effect of the number of nodes on energy consumption, the number of nodes involved in tracking in terms of communication range, and losing the target. This means the proposed method is superior and more efficient.

TABLE 5. THE RELATIONSHIP BETWEEN DESIGNING CRITERIA AND QUALITY TRACKING METRICS

| Layers and Services | Tracking Quality Criteria |
|---|---|
| Request, Routing of Integration Strategies, and Data Aggregation | Scalability of the Algorithm<br>Tracking Accuracy<br>Tracking Delay<br>Energy Consumption of Nodes<br>Control Packets Overhead |
| Target Mobility Model | Scalability of the Algorithm<br>Energy Consumption of Nodes<br>Flexibility of the Algorithm |
| Node Positioning Model | Tracking Accuracy<br>Energy Consumption of Nodes<br>Area Coverage<br>Control Packets Overhead |
| Locating Algorithm | Tracking Accuracy<br>Energy Consumption of Nodes<br>Control Packets Overhead |
| Scheduling Algorithm | Flexibility of the Algorithm<br>Tracking Accuracy<br>Area Coverage<br>Energy Consumption of Nodes<br>Control Packets Overhead |
| MAC Protocols | Scalability of the Algorithm<br>Energy Consumption of Nodes<br>Control Packets Overhead |

Reducing energy consumption and optimization are two important topics in tracking. That is because the nodes are scattered in the network and it is hard to access them for battery replacement. Therefore, the energy-centric protocol in the network should be designed carefully. As mentioned above, sensors with IDs and time slots have been used to implement the tracking protocol. In the proposed algorithm, scalability and energy consumption reduction for WSNs was considered in tracking. In this method, the detector nodes determine the location of the target using locating methods and locations of detector nodes. To reduce energy consumption at any given moment, the least number of sensor nodes were activated in the tracking area. The proposed method was implemented and compared with the basic method of EATT [22] through simulation. As discussed, the proposed method has a better energy consumption than the basic method. In addition, the proposed method had high throughput due to using time slots and this evaluation criterion has been investigated in number of iterations. The number of bits sent by the node and the number of bits reached in the network (with different rates) were reviewed. The high throughput of the proposed method is due to the use of time slots. This paper only focuses on movement of the target in one rout and its exit from the environment. In order to enhance the accuracy of prediction of the target movement, the records of each target can be kept, so that they can be used for better tracking of the target the next time it returns to the environment. In this regard, two type of environment can be examined: environments where one type of target enters and environment where several types of target enters. In the later environments, the target type is identified and it is recognized based on its acceleration, speed, and direction. Based on the proposed method in this study, after identification of the target, a more accurate prediction with fewer errors can be made considering the previous information about the target pattern.

Future works are summarized as follows:
- Changing the medium access layer and reassessing target tracking
- Examining metaheuristic methods, such as ant and PSO algorithms for improving target tracking
- Examining delays in the presented working scenario
- Using time slots considering the Markov forecast algorithm

**ACKNOWLEDGMENTS**

This work is supported in part by the National Natural Science Foundation of China under Grants 61632009 & 61472451, in part by the Guangdong Provincial Natural Science Foundation under Grant 2017A030308006 and Hgh-Level Talents Program of Higher Education in Guangdong Province under Grant 2016ZJ01.